# Labeling of Query Words using Conditional Random Field

Satanu Ghosh
West Bengal University of Technology
+91-7278137003
satanu.ghosh.94@gmail.com

Souvick Ghosh
Jadavpur University
+91-9007728924
souvick.gh@gmail.com

Dipankar Das
Jadavpur University
+91-9432226464
dipankar.dipnil2005@gmail.com

## ABSTRACT
This paper describes our approach on Query Word Labeling as an attempt in the shared task on Mixed Script Information Retrieval at Forum for Information Retrieval Evaluation (FIRE) 2015. The query is written in Roman script and the words were in English or transliterated from Indian regional languages. A total of eight Indian languages were present in addition to English. We also identified the Named Entities and special symbols as part of our task. A CRF based machine learning framework was used for labeling the individual words with their corresponding language labels. We used a dictionary based approach for language identification. We also took into account the context of the word while identifying the language. Our system demonstrated an overall accuracy of 75.5% for token level language identification. The strict F-measure scores for the identification of token level language labels for Bengali, English and Hindi are 0.7486, 0.892 and 0.7972 respectively. The overall weighted F-measure of our system was 0.7498.

## CCS Concepts
• **Computing methodologies~Natural language processing**
• **Computing methodologies~Information extraction**

## 1. INTRODUCTION
Language Identification is a necessary prerequisite for processing any user generated text, where the language is unknown. The identification of the language can be done at document level or at word level.

While Language Identification was previously being considered as a solved problem, the recent proliferation of social media and various phenomena such as code-switching, code-mixing, lexical borrowings and phonetic typing have introduced a new dimension to the problem. Random contractions (''em' in place of 'them' or 'shan't' in place of 'shall not') and transliterations have further complicated the problem of Language Identification. Various spelling variations, transliterations and non-adherence to formal grammar are also quite common in such text. [11, 14]

Language Identification for documents is a well-studied natural language problem [2]. King and Abney [6] presented the different aspects of this problem and focused on the problem of labeling the language of individual words in a set of multilingual document. They proposed language identification at the word level in mixed language documents instead of sentence level identification. The last few decades have seen the development of transliteration systems for Asian languages. Some notable transliteration systems were built for Chinese [7], Japanese [4], Korean [5], Arabic [1], etc. Transliteration systems were also developed for Indian languages [3, 9].

## 2. TASK DEFINITION
A query q : < $w_1w_2w_3 ... w_n$ > is written in Roman script. The words, $w_1,w_2,w_3, ... w_n$, could be standard English words or transliterated from Indian languages (L). The languages (L) can be Bengali (Bn), English (En), Gujarati (Gu), Hindi (Hi), Kannada (Ka), Malayalam (Ml), Marathi (Mr), Tamil (Ta) or Telugu (Te). The objective of the task is to identify the words as English or member of L depending on whether it is a standard English word or a transliterated L-language word. The words of a single query usually come from 1 or 2 languages and very rarely from 3 languages. In case of mixed language queries, one of the languages is either English or Hindi. Thus, queries are formed by mixing Tamil and English words, or Bengali and Hindi words, but not for example, Gujarati and Kannada words. We were also required to identify the Named Entities as NE (e.g. Sachin Tendulkar, Kolkata, etc).

## 3. DATASET AND RESOURCES
This section describes the dataset that have been used in this work. The training and the test data have been constructed using manual and automated techniques and made available to the task participants by the organizers. The training dataset consists of 2908 sentences whereas the test set contains 792 sentences.

The following resources provided by the organizers were also employed:

- **English word frequency list**[1]: It is in plain tab-separated text file containing English words collected from standard dictionary and followed by their frequencies computed from a large corpus. It contains noise (very low frequency entries) as it is constructed from news corpora.

- **Hindi word transliteration pairs 1** [10]: It is in plain tab-separated text file containing a total of 30,823 transliterated Hindi words (in Roman script) followed by the same word in Devanagari. It also contains Roman spelling variations for the same Hindi words (the transliteration pairs found using alignment of Bollywood song lyrics). However, it does not contain frequency or occurrence of a particular word transliteration pair.

- **Bangla word frequency list**[2]: It is in plain tab-separated text format. It contains Bengali words (Roman

---
[1] http://cse.iitkgp.ac.in/resgrp/cnerg/qa/fire13translit/index.html

[2] http://cse.iitkgp.ac.in/resgrp/cnerg/qa/fire13translit/index.html

script, ITRANS format) followed by their frequency computed from a large Anandabazar Patrika news corpus. ITRANS to UTF-8 converter is used for obtaining the words in Bengali script.

- **Gujarati word transliteration pairs**[2]: It is in plain tab-separated text format. It contains transliterated Gujarati words (Roman script) followed by the same word in Gujarati script. Due to the poor availability of Gujarati resources, a small list of 546 entries was created from training the data of FIRE shared task.

- **Google Input Tools**[3]: We used the lookup table of transliterated word pairs provided in Google Input Tools. These contain transliterated pairs of native Indian languages to Roman Script. We used these tables for all 8 Indian languages to create word list for each language.

- **Corncob Web Dictionary**[4]: The dictionary contains 58110 distinct English words. We have used it to identify English words.

- **Stanford NE Tagger**[5]: Named Entity Recognition (NER) labels sequences of words in a text which are the names of things, such as person and company names, or gene and protein names, etc.

We also developed 11 lists of our own which are as follows:

- *Named Entity List*: We developed this named entity list using the training data. It contains 648 distinct names.

- *Emoticon List*: We developed this list using Wikipedia. This list contains 273 distinct emoticons.

- *Language Wordlist*: We developed nine wordlists for nine different languages using training data. The wordlists contained few overlapping words.

## 4. SYSTEM DESCRIPTION
Our primary task was word-level language classification. However, identification of Named Entities was also necessary.

## 4.1 Word-level Language Identification Features
The following features were used for language identification:

### 4.1.1 Capitalization
Three types of Boolean capitalization features are used for encoding capitalization information. As all the words are in Roman script we use the ASCII value to identify a capital character. The first feature is whether the first character of the word is capital or not. This is an important feature as this is later used for identification of Named Entity. The second feature is whether the whole word is capital or not. The third feature is if any character in the word is capital or not.
For example, words like Mumbai, BCSE, 3G, etc.
CAP1: Is first letter capitalized? If yes, then CAP1 = 1, else 0
CAP2: Is any character capitalized? If yes, then CAP2=1, else 0
CAP3: Are all characters capitalized? If yes, then CAP3=1, else 0

---
[3] https://www.google.com/inputtools/

[4] http://www.mieliestronk.com/wordlist.html

[5] http://nlp.stanford.edu/software/CRF-NER.shtml

### 4.1.2 Word-level Context
The previous three words and the next three words along with the current token and length of the current token is used as contextual feature. As language identification and points of code-switch are context sensitive [12, 18, 19] we have used this feature only for classification. This feature is very much crucial to resolve the ambiguity in the word-level language identification problem. Let us consider examples given below:
- Mama *take* this badge off of me.
- Ami *take* boli je ami bansdronir kichu agei thaki.

The word `take' exists in the English vocabulary. However, the backward transliteration of `take' is a valid Bengali word. Words like `take', `are', `pore', and `bad' are truly ambiguous words with respect to the word-level language identification problem as they are valid English words as well as their backward transliterations are valid Bengali words. In this regard, context of the word can be used to correctly identify the language for such an ambiguous word. The dynamic unigram feature in the CRF++ template file analyses the previous token and the next token for their language and the language of the current token is annotated according to the context. Therefore, we have considered it as a very useful feature.
CON1: Current token
CON2: Previous 3 and next 3 tokens
CON3: Length of the current token. This feature is important because words in Indian languages tend to be longer than words in English.

### 4.1.3 Special Character
A word might start with some symbol, e.g. #, @, etc. These boolean features indicate the presence of hashtag (#), at the rate (@), hyperlink and emoticons. A list of emoticons containing 273 distinct emoticons using different kind of special characters was made and used for identification of emoticons.
For example, @aapyogendra, #aapsweep, http://t.co/pym4cr6xx0, :/
CHR1: If the word starts with #? If yes, then 1 else 0
CHR2: If the word starts with @? If yes, then 1 else 0
CHR3: If the word starts with http? If yes, then 1 else 0
CHR4: If emoticon? If yes, then 1 else 0

### 4.1.4 Dictionary Feature
A total of 9 different languages were there to be identified. We used 9 different lexical resources, one for each language. We used 9 different Boolean features to represent if a particular token is present in a particular lexicon. If a particular word is present in more than one lexicon, we use a unigram relational feature in the template file of CRF++ to handle the ambiguity. This unigram relational feature is determined using two or more other features.
For example, U1: %x[0,20]/%x[0,21]
LEX1: Is present in English dictionary? If yes, then 1, else 0
LEX2, LEX3,,…, LEX9 for other languages.

### 4.1.5 Presence of Symbol in word
Only one Boolean feature is used to identify the words with punctuation marks present in it. The punctuation marks can be an apostrophe ('), a dash (-), etc.
For example, goalkeepers\, angul-er
CHR5: Is symbol present? If yes, then 1 else 0

### 4.1.6 Presence of Digit
This Boolean function is used to indicate if a word contains a digit. As the corpus provided contains social media text, this feature was used. In phonetic script people often use digit to shorten their text.
For example 'gr8' in place of 'great', '4nds' for 'friends'
CHR6: Is digit present? If yes, then 1 else 0

### 4.1.7 Number Identification
This Boolean feature is used to identify if the token is number or not. For example, number like 30, 67, etc.
CHR7: Is token a number? If yes, then 1 else 0

### 4.1.8 Named Entity Identification
For NE identification we use the Stanford NE Tagger[6] along with a lexicon of named entities. We use two Boolean features for this purpose. The first is the basic lexicon search and the second is for the Stanford NE Tagger. We use another unigram relational feature in CRF++ for classification of NE Tags. The basic lexicon is the Named Entity list which we developed for our task.
NE1: If name entity matches List1, then NE1 = 1, else 0
NE2: If name entity matches List2, then NE2 = 1, else 0

## 5. RESULTS
In this work, Conditional Random Field (CRF) [13] has been used to build the framework for word-level language identification classifier. We have used CRF++ toolkit[7] which is a simple, customizable, and open source implementation of CRF.

The accuracies with respect to nine different languages as well as average and weighted F-measures are shown in Table 1 and Table 2.

**Table 1: Tokens level Results for language identification**

| Language | Precision | Recall | F-Measure |
|---|---|---|---|
| X | 0.9423 | 0.7525 | 0.8367 |
| Bengali | 0.8129 | 0.6937 | 0.7486 |
| English | 0.9318 | 0.8555 | 0.892 |
| Gujarati | 0.0757 | 0.4118 | 0.1279 |
| Hindi | 0.7772 | 0.8182 | 0.7972 |
| Kannada | 0.2793 | 0.799 | 0.4139 |
| Malayalam | 0.2597 | 0.6522 | 0.3715 |
| Marathi | 0.4956 | 0.8687 | 0.6311 |
| Tamil | 0.5672 | 0.817 | 0.6696 |
| Telegu | 0.3874 | 0.8153 | 0.5252 |

**Table 2: Other performance metrics**

| | |
|---|---|
| Tokens Accuracy (in %) | 75.4896 |
| Utterances Accuracy (in %) | 21.5909 |
| Average F-Measure | 0.538392 |
| Weighted F-Measure | 0.749833 |

**Table 3: Confusion matrix between languages**

| | en | X | hi | bn | ml | mr | kn | te | gu | ta |
|---|---|---|---|---|---|---|---|---|---|---|
| en | 3772 | 79 | 37 | 47 | 1 | 2 | 1 | 16 | 1 | 6 |
| X | 32 | 1763 | 2 | 1 | 0 | 0 | 0 | 1 | 0 | 0 |
| hi | 141 | 84 | 1242 | 38 | 0 | 6 | 3 | 6 | 9 | 0 |
| bn | 84 | 71 | 50 | 1112 | 0 | 7 | 2 | 4 | 9 | 8 |
| ml | 19 | 38 | 2 | 13 | 60 | 1 | 12 | 0 | 0 | 13 |
| mr | 23 | 33 | 53 | 65 | 2 | 225 | 3 | 2 | 1 | 1 |
| kn | 59 | 93 | 8 | 109 | 2 | 2 | 167 | 10 | 0 | 19 |
| te | 54 | 50 | 22 | 102 | 5 | 9 | 5 | 203 | 0 | 6 |
| gu | 18 | 13 | 77 | 39 | 0 | 3 | 6 | 0 | 14 | 9 |
| ta | 33 | 74 | 3 | 4 | 20 | 0 | 5 | 0 | 0 | 308 |

## 6. ERROR ANALYSIS
If we look at the confusion matrix for different languages, we can notice that many other languages have been wrongly classified as English. This is primarily due to overlapping words between English and all other Indian languages. In our task, the accuracies of MIXes and NEs were quite low. The primary reason for the increased error rate in MIX determination was the absence of post processing measures to identify the mixed words. Also the sub-classification errors in NE recognition could have been significantly reduced by adding a NE-classification module to our system. Our accuracy also declined for Gujarati, Kannada and Malayalam. Use of larger wordlists and transliterated dictionary should have improved the scores.

## 7. CONCLUSION
In this paper, we presented a brief overview of our system to address the automatic identification of word-level language. While the CRF-based approach was satisfactory, the results could have been improved by including post-processing heuristics for identifying mixed words and named entities. Using more character level features should improve the accuracy of the system. Also some basic knowledge about other languages and better wordlists and dictionary for regional languages should improve the accuracy of the present system. We used character n-grams (n=1 to 5) as one of the features of CRF++. However, the performance of the system declined on incorporating it.

## 8. ACKNOWLEDGMENTS
Our thanks to the organizers of FIRE 2015 shared task. Royal Sequira of Microsoft Research was very helpful throughout the work. We would also like to thank Nagesh Bhattu who corrected the annotations of numeric entities in the training data.

---

[6] http://nlp.stanford.edu/software/CRF-NER.shtml

[7] http://crfpp.googlecode.com/svn/trunk/doc/index.html